\documentclass[letterpaper,twocolumn,10pt]{article}
\usepackage{usenix-2020-09}

\usepackage{tikz}
\usepackage{amsmath}

\usepackage{times,url,color,soul,xspace,enumitem}
\usepackage{graphicx}
\usepackage{caption}
\usepackage{subcaption}
\usepackage{comment}
\usepackage{xspace}
\usepackage{tikz}
\usepackage{amsmath}
\usepackage[inline,draft,nomargin,index]{fixme}
\usepackage{footmisc}
\usepackage[frozencache=true,cachedir=./minted-main/]{minted}
\usepackage{adjustbox}
\usepackage{listings}

\usepackage{cleveref}
\crefformat{section}{\S#2#1#3} 
\crefformat{subsection}{\S#2#1#3}
\crefformat{subsubsection}{\S#2#1#3}

\FXRegisterAuthor{giu}{agiu}{\textcolor{red}{Giu}}

\FXRegisterAuthor{atr}{aatr}{\textcolor{blue}{Atr}}

\FXRegisterAuthor{lin}{alin}{\textcolor{green}{Lin}}

\setminted{fontsize=\footnotesize}

\newcommand{\unsim}{\mathord{\sim}}
\newcommand{\pangraph}[1]{\vspace{-5pt}\paragraph{#1}}

\makeatletter
\g@addto@macro{\UrlBreaks}{%
\do\/\do\a\do\b\do\c\do\d\do\e\do\f%
\do\g\do\h\do\i\do\j\do\k\do\l\do\m%
\do\n\do\o\do\p\do\q\do\r\do\s\do\t%
\do\u\do\v\do\w\do\x\do\y\do\z%
\do\A\do\B\do\C\do\D\do\E\do\F\do\G%
\do\H\do\I\do\J\do\K\do\L\do\M\do\N%
\do\O\do\P\do\Q\do\R\do\S\do\T\do\U%
\do\V\do\W\do\X\do\Y\do\Z}
\makeatother

\begin{document}

\title{\Large \bf Bento and the Art of Repeated Research}

\author{
{\rm Peter-Jan Gootzen and Animesh Trivedi}\\
VU University, Amsterdam\\
August 23, 2021
} 

\maketitle

\begin{abstract}
Bento provides a new approach to developing filesystems, with safety and high-velocity development in mind. This is achieved by using Rust, a modern and memory-safe systems programming language, and by providing a framework to run a single filesystem implementation in kernel space with VFS or in user space with FUSE.

In this paper, the benchmarking experiments from the Bento paper are repeated. We fail to exactly reproduce the results of the Bento paper, but more or less find the same patterns albeit with more outlying results.
Additionally we unsuccessfully run a standardized test suite, and expand the set of experiments with latency benchmarks and throughput benchmarks using a RAM block device. The latency benchmarks show that ext4 with journaling consistently outperforms Bento-fs and the RAM throughput benchmarks show no additional consistent performance pattern. During this experimentation, a set of 12 bugs was encountered and analyzed. We find that the ratio of memory related bugs is lower than other systems programming projects that use C as opposed to Rust.
\end{abstract}

\section{Introduction}

Filesystems is one of the systems programming areas that has the need for reliability combined with being per definition very low-level and "unsafe". The data stored with filesystems can be of high value, but a single bit flipped too much can lead to total corruption. Because of this filesystem research has not only focused on performance, but also on safety and reliability through approaches such as model checking \cite{explode}, fuzzing \cite{syzkaller} and record-and-play frameworks \cite{crashmonkey, AllFileSystemsAreNotCreatedEqual}.\smallskip

The C programming language is almost 50 years old, and has undergone several standardization revisions with most recently C17 from 2018. But at it's core, the language has stayed mostly the same, a high-level "portable assembler"\cite{C-portable-assembler}. To this day it is overwhelmingly still the most popular choice for systems programming, most notably in all kernel-related software. Fundamentally C is an unsafe language, memory related bugs are easily introduced and the various compilers do little to prevent this. Multiple large scale systems programming projects found that around 70\% of all bugs were memory related \cite{google-memory, microsoft-memory, bento}.

Rust is a relatively young programming language that is aimed at systems programming while being a more modern programming language with many features lent from functional languages and a proper dependency system. Its big differentiator however is that it allows for much safer systems programming with no extra cost nor overhead compared to C. It does this by employing a memory model that keeps track of the lifetimes of memory, which can be traced and verified for correctness by the compiler. Essentially creating an almost "compile-time garbage collector", giving the safety advantage that garbage collectors provide whilst simultaneously having the predictable performance that comes with manual memory management\cite{rustonomicon}. This memory model falls under the term "zero-cost abstractions" that was popularized by C++ creator Bjarne Stroustrup, and should therefore incur little to no overhead.

\subsection{Bento \& Rust}
At \textit{19th USENIX Conference on File and Storage Technologies} the paper \textit{High Velocity Kernel File Systems with Bento} was published that brings the powers of Rust to the systems programming paradigm of filesystems~\cite{bento}. Rust is a very suitable language for filesystems as it allows for the unsafe but necessary low-level code to be contained and thus easily verified, while the all the other code can be statically verified for memory correctness by the Rust compiler.

Apart from the earlier mentioned powerful memory model, Rust also provides many more features that make it a more modern and safe systems programming language. Three features of Rust that Bento employs will be further expanded upon in this subsection.

\pangraph{Enums:}
Enums in Rust are more akin to \textit{algebraic data types} from functional programming languages than enums from C-like languages, and also similar to the \texttt{union} type of C. A variant of an enum in Rust can hold data. The most prominent and clear example of this is \mintinline{rust}|enum Results<T, E>| (see Listing \ref{lis:result}), that can either be an \mintinline{rust}|Ok(T)| with a value of type \mintinline{rust}|T| on success or an \mintinline{rust}|Err(E)| with a value of type \mintinline{rust}|E| on failure. When the programmer receives this enum from a function call, it has to either handle both cases or explicitly let the program crash if it is not the expected value\cite{rust-book}.
This provides a safer approach to error handling than what is most commonly used in C with an integer return value and an \textit{out}-pointer parameter that the compiler cannot verify.
\begin{listing}[h!]
\begin{minted}{rust}
pub enum Result<T, E> {
    Ok(T),
    Err(E),
}
\end{minted}
\caption{Source code (simplified) of \mintinline[fontsize=\footnotesize]{rust}|enum Results<T, E>|}
\label{lis:result}
\end{listing}

\pangraph{Cargo:}
Using external dependencies in a C-based project is far from a simple process. Build-systems vary widely across codebases, function signatures in C-APIs are often very non-descriptive and reliance on compiler specific behavior is often a hurdle \cite{linux-clang}.

The Rust compiler \textit{rustc} provides a \textit{gcc}-like command-line interface, but the main way to interact with the Rust compiler is through the \textit{cargo} package manager and build system. Through \textit{cargo} dependencies from \url{https://crates.io} and git repositories can be included in a Rust project. This allows developers to easily include pieces of code that can be more easily verified for safety as only the \mintinline{rust}|unsafe| blocks need to be verified for correctness.

Bento-fs makes use of 79 dependencies (including dependencies of dependencies), ranging from a hash function implementation, to a library providing convenient logging macros to a library providing bindings to Linux kernel APIs.

\pangraph{Concurrency:}
To achieve optimal performance in any facet of systems programming, multi-threading and by extension concurrency, must be employed. However concurrency is an unsafe construct in C, as data races and undefined behavior can easily occur. Rust solves this by making its typesystem concurrency-aware through the \mintinline{rust}|Send| and \mintinline{rust}|Sync| types, that specify whether sole ownership of an object is safe to transfer to another thread and whether access from multiple threads to a single object is safe, respectively \cite{rustonomicon}. This furthermore increases the ease of including external dependencies as misuse of a library that employs concurrency will be caught by the compiler.

Bento and Bento-fs make extensive use of this feature of the Rust typesystem as filesystems are inherently very concurrent. An interesting use of this is the wrapper interface that is employed to use the Linux kernel \textit{semaphore} type in safe and idiomatic Rust.

\subsection{Repeated research on Bento}
In this paper we will repeat the experiments from the Bento paper and attempt to reproduce the numbers from the original paper (\autoref{sec:reproduction}). This will be followed up with additional latency benchmarks (\autoref{sec:latency}), and throughput benchmarks using a RAM-backed block device (\autoref{sec:ram}). These additional benchmarks are hypothesized to show additional performance discrepancies between Bento-fs (the ext4-like filesystem with journaling implemented on top of the Bento framework) and ext4 with journaling. To conclude the experimentation, a standardized filesystem test suite will be ran on Bento-fs (\autoref{sec:testsuite}). 

\begin{table*}[h]
    \begin{minipage}[t]{1.0\textwidth}
        \begin{adjustbox}{width=\columnwidth,center}
            \begin{tabular}{|l|l|l|l|l|l|l|}
            \hline
            No & Kernel & FS         & Cause                                       & Module  & Function                              & Symptom               \\ \hline
            1  & 4.15   & Bento-fs   & \textit{createfiles}/\textit{deletefiles} rm                  & ?       & ?                                     & syscall hang          \\ \hline
            2  & 4.15   & Bento-fs   & repeated \textit{gitclone} after other experiments   & bentofs & bento\_file\_put \& bento\_xattr\_set & deadlock              \\ \hline
            3  & 4.15   & Bento-fs   & repeated \textit{tar}/\textit{untar} after other experiments  & ?       & ?                                     & folder disappears     \\ \hline
            4  & 4.15   & Bento-fs   & repeated \textit{webserver} or Filebench seq read              & ?       & jbd2 commit                           & soft lockup           \\ \hline
            5  & 4.15   & Bento-fs   & umount after repeated Filebench             & bentofs & bento\_flush\_times                   & invalid page request  \\ \hline
            6  & 4.15   & Bento-fs   & repeated seq write or xfstests \textit{generic/013}  & bentofs & bento\_file\_put                      & deadlock              \\ \hline
            7  & 4.15   & Bento-fs   & ?                                           & xv6fs   & radix\_tree\_lookup\_slot             & missing page cache    \\ \hline
            8  & 4.15   & Bento-prov & \textit{deletefiles} with T=18 and unmount                    & bentofs & bento\_do\_getattr                    & invalid page request  \\ \hline
            9  & 4.15   & Bento-user & \textit{tar}                                         & ?       & ?                                     & I/O error             \\ \hline
            10 & 4.15   & Bento-fs   & random read fio                                & ?       & ?                                     & I/O error             \\ \hline
            11 & 4.15   & Bento-fs   & \textit{rocksdb} with folder in root of disk remount & xv6fs   & bento::kernel::journal::begin\_op     & jbd2 too many credits \\ \hline
            12 & 5.4    & Bento-user & \textit{createfiles}, then remount, then rerun       & xv6fs   & bento::fuse::request::dispatch        & Invalid FUSE opcode   \\ \hline
            \end{tabular}
        \end{adjustbox}
        \captionsetup{justification=centering,margin=1cm}
        \caption{Overview and summary of the encountered bugs with Bento. "*" in the Kernel column signifies that it occurs on all supported kernel versions and "?" signifies that this was unknown or unable to be determined.}
        \label{tab:bugs}
    \end{minipage}
\end{table*}

\section{Bugs, bugs, bugs}\label{sec:bugs}
The experiments are executed under slightly different circumstances than in the original paper (e.g. different order of tests), as the experiment suite is recreated from scratch. This together with previously not ran experiments, results in the finding of 12 bugs that are shown in \autoref{tab:bugs}. This section discusses the bugs that are found under the Bento version that targets Linux 4.15, the bugs found with the Linux 5.4 version, the mitigations taken to still perform the experiments and lastly an analysis of the bugs is performed to investigate the nature of the bugs and how they relate to Rust.

\subsection{Bugs encountered on Linux 4.15}
The first bug does not cause any crash, but it makes the filesystem unusable. The \textit{createfiles} and \textit{deletefiles} workloads create a very large \textit{fileset} with 800 thousand files and 55 thousand files in a directory, respectively. Every time these workloads are ran, the fileset from the previous run is completely deleted. But the deletion of this large \textit{fileset} becomes very slow. Specifically the \textit{newfsstat} system call that the \textit{rm} commands executes, becomes slow, in the case of the \textit{createfiles} workload, it does not return after a full hour. This indicates that some part of the Bento code's complexity does not scale as well as ext4. 

The second bug occurs when the system is under repeated stress from multiple benchmarks. When the second set of application workloads is ran with the \texttt{git clone} experiment after the others, a deadlock occurs between a kernel thread and a Bento thread that is reported by the kernel in the kernel ring buffer (i.e. the \textit{dmesg} logs).
When this set of application workloads is ran with \texttt{tar} and \texttt{untar} last, the third bug is triggered, the folder where the results are outputted to will suddenly be "forgotten" by the filesystem with no crash or other noticeable side-effect.\\
\indent The fourth bug occurs when the \textit{webserver} Filebench workload is ran five times in a row, or occurs very infrequently during a sequential read Filebench workload run. This bug causes a jbd2 transaction commit to consistently soft lockup.
The fifth bug only occured once when unmounting a Bento-fs disk after running various Filebench workloads, where an invalid kernel page request is issued.
The sixth bug occurs for all sequential write Filebench workloads at run $>1$ (not consistent at which run) or when the \textit{generic/013} xfstests regression test is ran, and causes a pair of \textit{kworkers} to infinitely block.
The seventh bug concerned the sequential read workloads but was unable to be properly reproduced and caused a soft lockup.\\
\indent The eight bug occurred when the \textit{deletefiles} Filebench workload was ran with 18 threads on Bento-fs with provenance and subsequently the disk was unmounted, causing an invalid kernel page request.
The ninth encountered bug was a consistent I/O error during the \texttt{tar} benchmarks when mounting a Bento-fs formatted disk in user space.
The tenth bugs was also a consistent I/O error, but during the \textit{randread} fio benchmark with Bento-fs mounted in kernel space.\\
\indent The eleventh bug occurs when a disk is freshly imaged with Bento-fs, mounted via the kernel space module and the \textit{rocksdb} benchmark is ran inside of a folder in the root of the disk (e.g. /mnt/xv6fsll/rocksdb). This causes the filesystem to forget that the folder exists when a file is created in it, causing the benchmark to crash. When subsequently the filesystem is unmounted and remounted, the filesystem module crashes inside of the jbd2 Linux kernel module. 

\subsection{Bugs encountered on Linux 5.4}
The \textit{master} branch of the Bento framework targets Linux 4.15 and the experiments in the paper and in this paper were performed using Linux 4.15. The Bento repository contains a branch that seemingly supports the newer kernel version Linux 5.4. This code compiles and the kernel modules are able to be loaded into the kernel. Only one reproducible bug that wasn't already present on the original Linux 4.15 version was found. This deployment wasn't as stress tested as with Linux 4.15, so it likely that bugs were missed.

The twelfth and last bug of the 12, occurs when a fresh Bento-fs image is mounted via FUSE in user space, the \textit{createfiles} Filebench workload is ran, the image is unmounted, remounted in user space and finally when rerunning the workload, the filesystem kernel module calls an invalid FUSE call is made, thus causing a crash. 

\subsection{Bugs mitigations}
Luckily this set of bugs can be somewhat negated by only running the \textit{createfiles} and \textit{deletefiles} workloads once with a unique \textit{fileset} name, only running the \textit{webserver} workload four times, running the sequential write workloads only once and by running the set of application benchmarks in a certain order with only five runs and \textit{sleep} calls in between the benchmarks. The \texttt{untar} and \texttt{tar} benchmarks were unable to finish on Bento-fs in user space as a result of bug 9 and are therefore not included in the results. Lastly, because of bug 10, the \textit{randread} workload was completely omitted from the experiment.

This does not fully mitigate all bugs, as occasional I/O errors are still occurring (albeit very infrequently) and bug 4 still occurs very infrequently even with only one run of each workload. In this case benchmarks were simply reran.

\subsection{Bugs analysis}
As one of the main goals of Bento is to improve safety and reduce the amount of bugs, analyzing the bugs that were found might shed more light on whether these goals were met. Several of the bugs have no proper call trace and are therefore unable to be analyzed within the time frame of this research project. In total there were eight call traces collected that were reported in the kernel ring buffer.\\
\indent Of these eight bugs with a call trace, one (bug 4) was unable to be traced to a specific kernel module, but was in essence an infinite lockup during a jbd2 commit transaction. Likely being caused by improper use of the jbd2 API that is called from the filesystem kernel module written in Rust. Out of the other seven, four stem from the \textit{bentofs} kernel module that is written in C and acts as the connector between VFS and the filesystem implementing kernel module \footnote{Yes all these names are very confusing.}. Two of these (bug 2 and 6) call traces report a deadlock between a kworker and the application performing a syscall. The other two (bug 5 and 8) are invalid page requests that could have many causes like accessing already freed memory or overflowing a buffer. The \textit{bentofs} kernel module is written in C instead of Rust because this code is so fundamentally unsafe that programming it in Rust would not make sense as virtually all code would have to be marked as \mintinline{rust}|unsafe| to compile \cite{bento}. So although the borrow checker Rust prevents memory bugs, in this case unsafe code was inevitable, resulting in two bugs.\\
\indent The remaining three bugs can be traced to the Bento-fs or Bento-prov (i.e. the filesystem implementing) kernel modules called \textit{xv6fs}. The first of the three (bug 7) is a caused by a non existent page cache entry, that is assumed to be there, likely being a programming error (i.e. no edge case for when it isn't there). The second of the three (bug 12) occurs because a FUSE call is performed with an invalid FUSE opcode. This bug only occurs on the updated version of Bento that targets Linux 5.4, it is likely that the FUSE opcodes changed and that this is a regression bug.\\
\indent The last of these three and last of all the eight bugs with call traces (bug 11), is quite interesting because it has already been fixed after we reported the bug \footnote{\href{https://gitlab.cs.washington.edu/sm237/bento/-/commit/5f48bb55e8db51eaa300e06ffb018b2a9c01c69a}{commit that fixed bug 11}}. It occurred because the \mintinline{rust}|fn bento_destroy(&mut self, _req: &Request)| function, that the C-based \textit{bentofs} kernel module calls when it wants to unmount the Rust-based filesystem implementing kernel module, did not properly drop the reference to the jbd2 journal causing the old journal to be reused on remount. This bug could not be caught by the Rust typesystem because the filesystem object that encapsulates this journal reference, comes from a C-context as a static pointer which disallows for the typesystem to do the usual memory safety checks ("\textit{borrow checking}").\\
\indent Summing these findings up, we see that 3 out of 8 were C-related memory bugs that could have been prevented by Rust, but could not have been programmed in Rust, 2 were deadlocks that Rust would be unable to prevent and 3 were API misuses and programmer error bugs. Although this is a small sample of bugs, the percentage of memory bugs is found to be 37.5\%, roughly half of the commonly found $\unsim70\%$ in other systems programming projects (including a filesystem) \cite{google-memory, microsoft-memory, bento}.

\begin{figure*}[t]
	\centering
	\begin{minipage}[t]{.5\textwidth}
		\centering
		\includegraphics{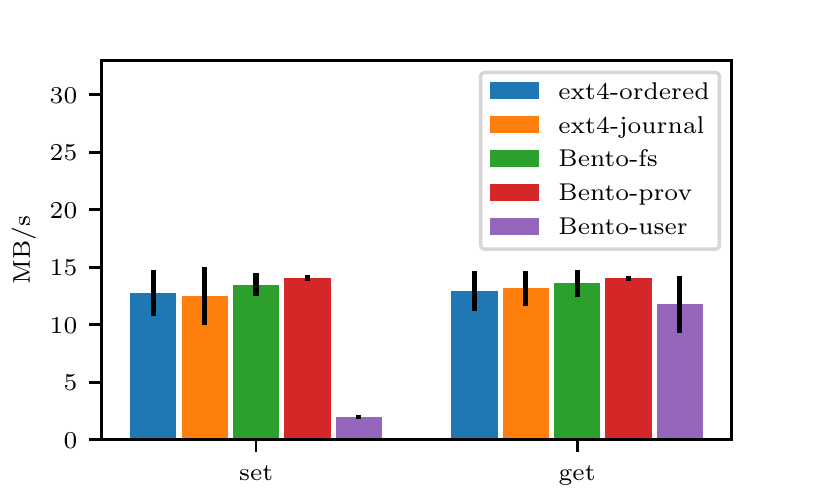}
		\captionsetup{justification=centering,margin=1cm}
		\caption{Results for Redis \textit{get} and \textit{set} benchmarks with Bento-fs, Bento-fs mounted through user space, Bento-fs with provenance and ext4 ( \texttt{data=journal}  and \texttt{data=ordered} )}
		\label{fig:redis}
	\end{minipage}\hfill
	\begin{minipage}[t]{.5\textwidth}
		\centering
		\includegraphics{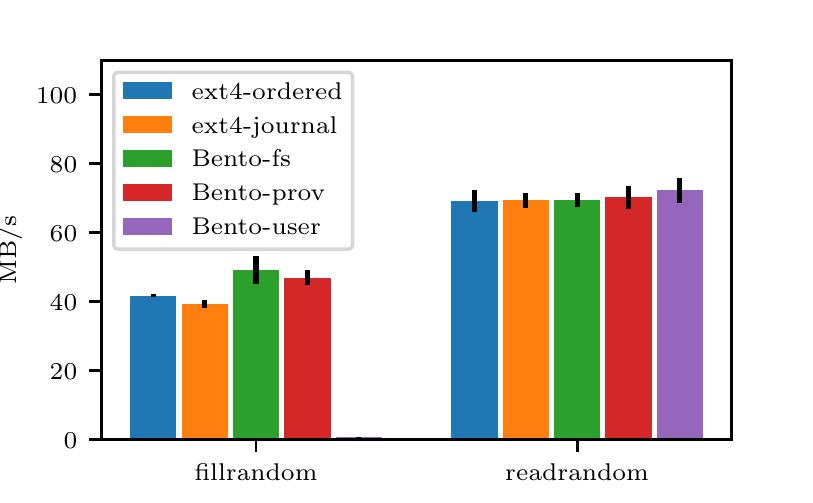}
		\captionsetup{justification=centering,margin=1cm}
		\caption{Results for Rocksdb \textit{fillrandom} and \textit{readrandom} benchmarks with Bento-fs, Bento-fs mounted through user space, Bento-fs with provenance and ext4 ( \texttt{data=journal}  and \texttt{data=ordered} )}
		\label{fig:rocksdb}
	\end{minipage}
\end{figure*}
\section{Experiments}\label{sec:experiments}
The Bento paper evaluates the performance of the implementation through several experiments that are ran on multiple versions of Bento-fs (kernel space, user space and provenance) and ext4 from the Linux kernel with and without journaling (i.e. \texttt{data=journal}  and \texttt{data=ordered} ). These experiments will be reproduced in this section. The reproduction of the live kernel module upgrade experiments will be skipped.

Subsequently in this section, the latency will be benchmarked with fio and a subset of the Filebench workloads from the Bento paper will be reran using a RAM block device.

The experiments in the paper were ran with two Intel Xeon Gold 6138 CPU's (2 sockets, each with 20 cores, 40 threads), 96 GB DDR4 RAM, and a 480 GB Intel Optane 900P SSD. We will run these same experiments inside of a QEMU KVM virtual machine on two Intel Xeon Silver 4210R CPU's (2 sockets, each with 10 cores, 20 threads), 256 GB DDR4 RAM, and a 240 GB Intel Optane 900P SSD (different capacity but the same specifications as the 480 GB variant). The VM will have 18 of the 20 cores allocated, have access to 96GB DDR4 RAM and the Optane device is passthroughed via VFIO.

\subsection{Bento paper experiments reproduction}\label{sec:reproduction}
The first set of experiments is a set of Filebench workloads, that in the paper are ran with 1 and 40 threads but we will run with 1 and 18 threads to account for the fewer amount of cores in our machine (thread count will be denoted with T). The second set of experiments consists of several application workloads, namely \texttt{tar}, \texttt{untar} and \texttt{grep} using the Linux codebase, and two benchmarks with \textit{Redis} and \textit{Rocksdb}.

The Bento paper also contains a \textit{git clone} experiment that clones the xv6 git repository\footnote{\url{https://github.com/mit-pdos/xv6-riscv}} and measures execution time. We experienced high variability in the download speed from the Github server to our server rack at the VU when cloning this repository, causing the standard deviation of the execution time to be higher than the average execution time. For the sake of reproducibility we therefore have not included this experiment in our results.

The create and delete Filebench workloads are not properly specified, by contacting the authors of Bento it was found that these two workloads stem from \footnote{\href{https://github.com/filebench/filebench/blob/22620e602cbbebad90c0bd041896ebccf70dbf5f/workloads/filemicro_createfiles.f}{filemicro\_createfiles.f}\label{foot:createfiles.f}} and \footnote{\href{https://github.com/filebench/filebench/blob/22620e602cbbebad90c0bd041896ebccf70dbf5f/workloads/filemicro_delete.f}{filemicro\_delete.f}} with the parameters changed to increase the workload size (see \hyperref[sec:availability]{Availability}).
The \texttt{tar}, \texttt{untar}  and \texttt{grep}  benchmarks were also not properly specified, so widely used '\textit{tar -czf}' and '\textit{tar -xzf}' were used for \texttt{tar}  and \texttt{untar}  respectively, and with \texttt{grep} the pattern "jbd2\_journal\_start" was searched.

For more explicit explanations of experiments the Bento paper should be consulted.

\subsubsection{Results}
\begin{figure*}[t]
	\centering
	\begin{minipage}[t]{.5\textwidth}
		\centering
		\includegraphics{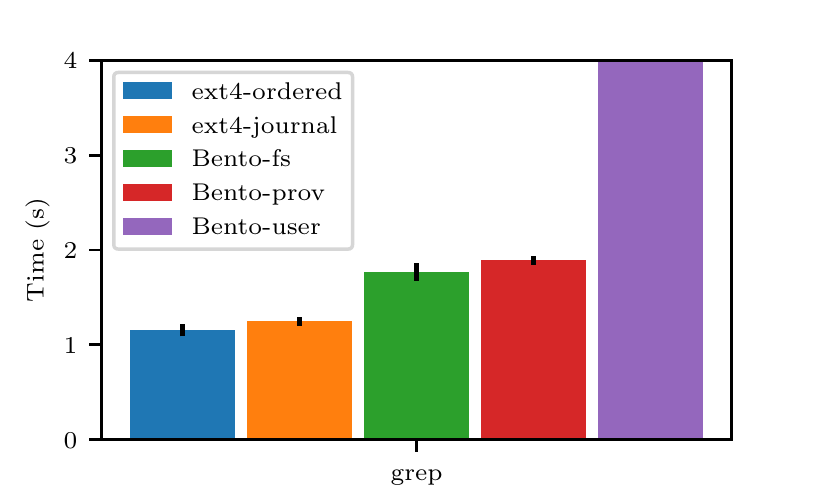}
		\captionsetup{justification=centering,margin=1cm}
		\caption{Results for \texttt{grep} benchmarks with Bento-fs, Bento-fs mounted through user space, Bento-fs with provenance and ext4 ( \texttt{data=journal}  and \texttt{data=ordered} ). The result for Bento-user is cut off, its average is 17.58 seconds and with a standard deviation of 3.83.}
		\label{fig:grep}
	\end{minipage}\hfill
	\begin{minipage}[t]{.5\textwidth}
		\centering
		\includegraphics{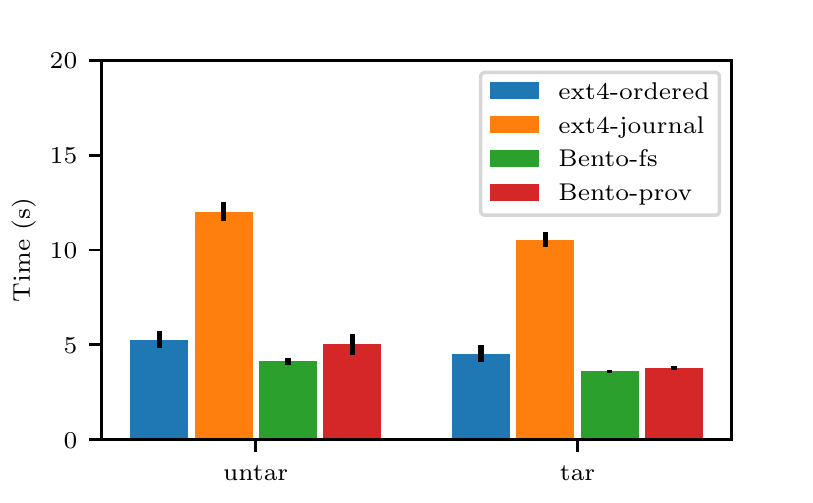}
		\captionsetup{justification=centering,margin=1cm}
		\caption{Results for \texttt{untar} and \texttt{tar} benchmarks with Bento-fs, Bento-fs mounted through user space, Bento-fs with provenance and ext4 ( \texttt{data=journal}  and \texttt{data=ordered} )}
		\label{fig:tar}
	\end{minipage}
\end{figure*}

\pangraph{Filebench workloads.}
In \autoref{tab:filebench} we see the raw results of the Filebench workloads. The first thing that stands out are the very high MB/s numbers reported by the read workloads, higher than the Intel Optane 900P can deliver (rated at 2.500MB/s read and 2.000MB/s write speeds). This is caused by caching within the kernel, as \textit{Direct I/O} is \textbf{not} used in these experiments.

Looking at the read speeds, we see that Bento-fs mounted through user space is able to very adequitely able to keep up with Bento-fs is kernel space, in some cases even slightly surpassing it by several hundred MB/s. Next looking at both ext4 versions, it is clear that Bento-fs in kernel space only outperforms them with single-threaded sequential reads, in all other ext4 fares better. The sequential read multithreading does not scale linearly, but Bento-fs and Bento-user scale worse (several GB/s worse) than the ext4's. The random read speeds are very similar across the different filesystems, indicating that the filesystems are equally bottlenecked by the 900P and caching in the kernel.

The write speeds paint a different picture, with Bento-fs through user space suffering from very poor write throughput performance compared to the kernel space filesystems, not even being able to surpass $15$MB/s. Even though Bento-fs does journaling, it performs more inline with the non-journaling counter-part of ext4, even surpassing it by several 100MB/s in most cases. The multithreading with random writes negatively affects the total throughput of all filesystems roughly equally.

\pangraph{Key-value stores.}
\autoref{fig:redis} and \autoref{fig:rocksdb} show the results for the key-value store benchmarks. The patterns found here are very similar to those found in the Filebench results. Firstly, Bento-fs (with and without provenance) is slightly more performant (less than $10\%$) than ext4. Secondly, the write benchmarks with Bento-user are much slower (one or more magnitudes of order) than with the kernel space filesystems. Thirdly and lastly, in the read benchmarks Bento-user almost equals the others with Redis and even performs a few percentage points better than the others with RocksDB.

\pangraph{Applications.}
In \autoref{fig:grep} and \autoref{fig:tar} the application benchmark results can be found. The \texttt{grep} benchmark is characterized by many file opens and reading through each file completely. The Bento filesystem with and without provenance show a significant decrease in performance compared to the ext4's. Bento-fs mounted through user space performs very poorly, with the running time increasing 10-fold and the variance increasing greatly.\\
\indent The \texttt{untar} and \texttt{tar} results show a different pattern, with Bento being faster than ext4, apart from ext4 with journaling showing significant $>2\times$ slowdown compared to its non-journaling counterpart. The performance overhead of provenance in Bento-fs is more significant with \texttt{untar}, this is easily explained by the fact that \texttt{untar} creates many new files triggering a provenance log, while \texttt{tar} only creates a single file.\\
\indent \autoref{fig:app_workloads} shows the results for the synthetic application workloads with Filebench. Once more we find that across the board, Bento-fs mounted through user space suffers from very poor performance compared to the other benchmarked filesystems (one or more orders of magnitude slower). These results show a significant performance disadvantage for Bento-fs across all workloads.

\subsubsection{Results comparison with Bento paper}
\begin{figure}[t]
	\centering
	\includegraphics{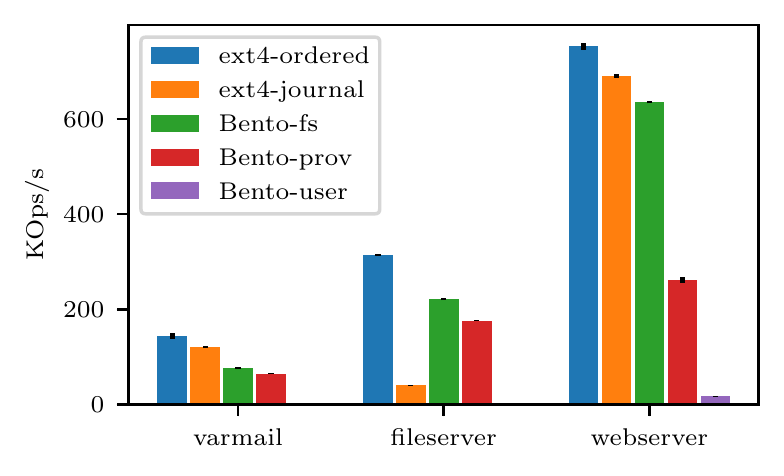}
	\caption{Results for \textit{webserver}, \textit{fileserver} and \textit{varmail} Filebench workloads with Bento-fs, Bento-fs mounted through user space, Bento-fs with provenance and ext4 (\texttt{data=journal} and \texttt{data=ordered})}
	\label{fig:app_workloads}
\end{figure}

When comparing the numbers from the experiments performed to those found in the Bento paper, it becomes immediately obvious that differences are very large, in some cases our results are even $10\times$ faster.
It is unclear why this is the case, as the hardware used is very similar and no known configuration changes have been made.
Because of this the patterns that were found in the Bento paper will be compared to the patterns found in our results, so as to account for this difference in absolute performance.\\
\indent \autoref{tab:filebench_cmp} shows a comparison of relative throughput performance found in our results and those found in the Bento paper. While in the Bento paper the measured read throughputs of Bento-fs and Bento-user were nearly equal to those of ext4-j (i.e. ext4 with \texttt{data=journal}), we find that the sequential read speeds differs significantly from ext4-j, trading blows depending on the workload. The write throughputs reported in the Bento paper indicate a performance increase for Bento-fs compared to ext4-j, we find this pattern too. This performance increase is for all workloads either significantly higher or lower than reported in the Bento paper, with no particular pattern. The \textit{createfiles} and \textit{deletefiles} workload results show roughly the same pattern as those reported in the paper (taking into account the thread count difference), except for the multithreaded \textit{createfiles}, which performs much poorer with Bento-fs compared to ext4-j in our results than in the original paper.\\
\indent Comparing the Filebench synthetic application workloads with those found in the paper (\autoref{fig:app_workloads} and Figure 2 \cite{bento}), we see that the same pattern appears with worse relative performance of Bento-fs with provenance reported in our numbers. 
The performance pattern found with the \texttt{grep} experiment in \autoref{fig:grep} compared to that found in the Bento paper (Figure 3 \cite{bento}), is the same but with more pronounced performance differences.
The key-value store benchmarks results from \autoref{fig:redis} and \autoref{fig:rocksdb} show the same pattern as Figure 4 in \cite{bento}, apart from small single digit percentage point deviations that let one filesystem slightly out edge another.

\begin{figure*}[t!]
	\centering
	\begin{subfigure}[t]{.33\textwidth}
		\centering
		\includegraphics{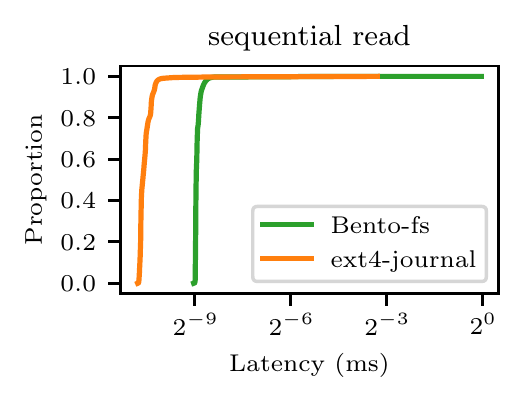}
	\end{subfigure}\hfill
	\begin{subfigure}[t]{.33\textwidth}
		\centering
		\includegraphics{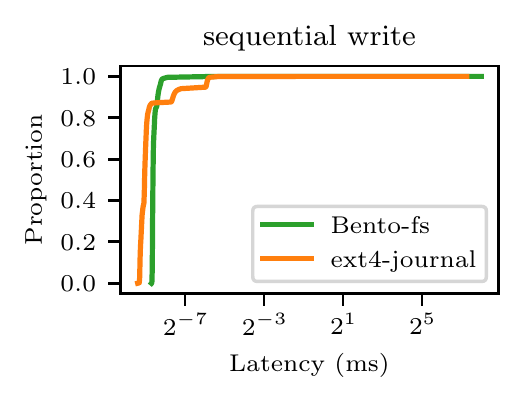}
	\end{subfigure}
	\begin{subfigure}[t]{.33\textwidth}
		\centering
		\includegraphics{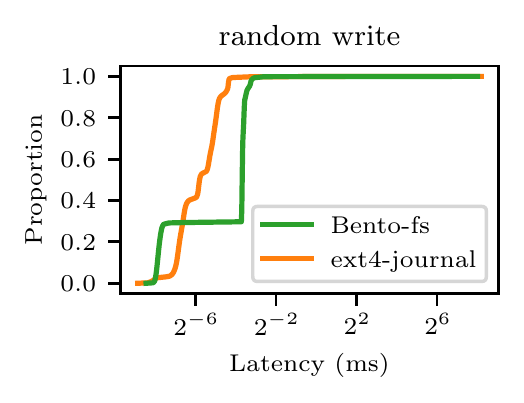}
	\end{subfigure}
	\captionsetup{justification=centering,margin=1cm}
	\caption{Latency measurement results with fio using three workload types with Bento-fs (kernel space) and ext4 with \texttt{data=journal}.}
	\label{fig:latency}
\end{figure*}

\subsection{Latency experiment}\label{sec:latency}
The experiments from the Bento paper focus on application-like benchmarks and synthetic throughput benchmarks. Although the application-like benchmarks also indirectly benchmark latency (albeit minimally), a separate synthetic latency experiment could provide valuable insight into the latency performance characteristics of the Bento and Bento-fs implementation.

\subsubsection{Latency experiment setup}
The synthetic latency experiment will be performed with fio by performing three types of operation workloads: sequential read, sequential write and random write. The random read workload is omitted from the experiment, as this workload crashes on Bento-fs (see \autoref{sec:bugs}). The three workloads each run for 5 seconds on a 4GB file, with a single thread, a 512B blocksize, an I/O depth of 1, the sync I/O engine and via Direct I/O to remove the kernel's caches from the picture. The experiment will be performed on Bento-fs (that is the kernel space version without provenance) and ext4 with journaling (as Bento-fs also performs journaling).

\subsubsection{Latency experiment results}
The three subfigures from \autoref{fig:latency} plot the results from the experiment. The first workload with sequential reads shows that ext4 has a clear latency advantage by roughly $4\times$, the difference increases in the $>99\%$ percentile to more than $8\times$. The sequential write workload shows a similar pattern with the $[0\%, 80\%]$ percentiles and the upper most percentile showing a slight latency advantage for ext4. However between $[80\%, 99\%]$ Bento-fs still provides latencies of the lower percentiles while ext4 has two performance knees, causing it to be slower than Bento-fs. The results of the random write workload show that neither filesystem has the complete upper hand, with Bento-fs producing lower latencies in the lowest 30th percentiles of latencies while ext4 overall has a lower average latency owing to the $[30\%, 100\%]$ percentile range.

\subsubsection{Latency discrepancy hypothesis}
With these results it is clear that overall, ext4 with journaling has better latency performance than Bento-fs. Even though the exact cause is unclear, two possible hypotheses come to mind when looking at the design of Bento.

The first being that Bento-fs runs on top of the Bento framework, i.e. each operation going through two kernel modules (the \textit{bentofs} kernel module that provides the framework and the filesystem implementing kernel module \textit{xv6fs} in the case of Bento-fs). This provides an extra level of indirection and thus extra latency that ext4 does not incur.

The second hypothesis is that because of the easy access to third-party dependencies and full-fledged standard library of Rust, extra latency overhead was incurred. Analysis of the Rust code would provide more insight into this hypothesis.

%
\subsection{RAM block device throughput benchmarks}\label{sec:ram}
The experiments in all previous subsections were performed on an Intel Optane 900P and were able to show a performance discrepancy between Bento-fs and ext4 with journaling. This experiment will run a subset of the Filebench throughput workloads with a RAM block device as the backing device. The theoretical bandwidth of the memory in our machine is 115.2GB/s, atleast 57 times more than what the Optane device can provide. It is hypothesized that this extra increase in available bandwidth will increase the discrepancy between the two filesystems. To test this hypothesis the random write, random read, sequential read and sequential write Filebench workloads will be ran with T=1 and T=18 on Bento-fs and ext4-journal.

\begin{table}[h]
    \centering
    \begin{tabular}{|l|r|r|}
        \hline
        \multicolumn{1}{|c|}{Workload} & \multicolumn{2}{c|}{Bento-fs:ext4-j} \\ \cline{2-3}
        \multicolumn{1}{|c|}{}         & \multicolumn{1}{c|}{900P} & \multicolumn{1}{c|}{RAM} \\ \hline
           seq\_read\_128k, T=1        &                      1.11 &                     0.86 \\ \hline
          seq\_read\_128k, T=18        &                      0.84 &                     0.84 \\ \hline
          rand\_read\_128k, T=1        &                      1.00 &                     1.00 \\ \hline
         rand\_read\_128k, T=18        &                      1.00 &                     1.00 \\ \hline
          seq\_write\_128k, T=1        &                      2.67 &                     2.69 \\ \hline
         rand\_write\_128k, T=1        &                      4.07 &                     3.99 \\ \hline
        rand\_write\_128k, T=18        &                      3.57 &                     4.06 \\ \hline
    \end{tabular}
    \captionsetup{justification=centering,margin=1cm}
    \caption{Results of the throughput RAM benchmarks shown as a comparison between Bento-fs mounted through kernel space and ext4 with journaling on both a RAM-backed block device and an Intel Optane 900P.}
    \label{tab:filebench_ram_cmp}
\end{table}

\subsubsection{Results}
In \autoref{tab:filebench_ram_cmp} the results of the RAM benchmarks are shown. To test the validity of the hypothesis, the throughput performance of Bento-fs compared to ext4-journal on a RAM block device is shown. Comparing these numbers to those found when using the Optane device, show that two workloads have significantly different ($>0.15$ difference) throughput performance characteristics on RAM compared to the Optane device. The single-threaded sequential read workload that performed better on Bento-fs with the Optane device, now performs better on ext4 by $\unsim900$MB/s, while the multi-threaded variant performs the same with RAM block device relative to ext4. The second workload that performs significantly different is the multithreaded random write workload, where Bento-fs gains $\unsim200$MB/s compared to ext4-j by using a RAM block device, compared to a $\unsim100$MB/s performance decrease with the Optane device.

As these results do not favor one filesystem and are inconclusive, we cannot conclude that a higher throughput capable device would increase the performance discrepancy between Bento-fs and ext4 with journaling.

\subsection{Validation through standardized filesystem test suite}\label{sec:testsuite}
Developing a production-ready filesystem is a difficult process. Filesystems have to fit into the complicated code and ecosystem of the kernel, perform low level operations to store and read data, and have to be very resilient to bugs because of the great value that will be stored on the disk with the filesystem.
To account for these complications, test suites have been developed to verify the validity of filesystem implementations. One of the most prominent filesytem test suites in the Linux-space is xfstests \cite{xfstests}. This test suite was originally developed for regression testing the XFS filesystem implementation in the Linux kernel, but was expanded to other filesystems such as NFS and Ext4. A portion of its tests are \textit{generic} (i.e. filesystem-agnostic) and can thus be used to validate any filesystem that targets Linux. Using this set of tests the Bento-fs filesystem will be validated.

\subsubsection{Accommodating xfstests for Bento-fs}
The Bento-fs filesystem has three mount options that are not-optional. The \textit{roodmode} option specifies the \textit{mode} of the root inode of the filesystem (Bento-fs always uses "40000") and \textit{user\_id} with \textit{group\_id} specify whom on the system is allowed to access the filesystem. These options were amended in the mount procedure of xfstests.

All tests require a \textit{DEV} disk that has to be preformatted with the to be tested filesystem. A subset of the tests require a second \textit{SCRATCH} disk that will be reformatted between tests and more frequently mounted and remounted. To run this subset of tests, a \textit{mkfs.{FILESYSTEM}} utility is needed that is complaint with the \textit{mkfs} command-line interface. Bento-fs however doesn't currently have such a \textit{mkfs.bentoblk} (the filesystem type of Bento-fs is \textit{bentoblk}) utility. The tests that require a \textit{SCRATCH} disk will therefore not be ran.

\subsubsection{Validation results}
When running the \textit{generic} group of xfstests (totaling 636 tests) on our machine, the first few tests either pass or are not supported by Bento-fs. However the \textit{generic/013} test already locks up (bug 6), which is reported by the kernel in the kernel ring buffer. This was to be expected with the many encountered bugs in \autoref{sec:reproduction}, \autoref{sec:latency} and \autoref{sec:ram}. No further effort will be done to perform standardized testing on Bento, because of this second confirmation that Bento is simply not stable.

\section{Conclusion}
We are unable to exactly reproduce the numbers from the experiments in the original Bento paper. However when analyzing the patterns found in our results and those found in the Bento paper, we find that most patterns match up with some slight deviations and more outlying results. Additional experiments show that in terms of latency, Bento-fs performs consistently worse than ext4 with journaling, and that the throughput performance of Bento-fs and ext4 with journaling scale equally well when using a RAM-backed block device.
We are therefore only able to partially back up the conclusion from the Bento paper, we find that Bento-fs has similar throughput and application performance to ext4 with journaling, but in terms of latency performance ext4 with journaling does outperform Bento-fs.
We find that Bento-fs does not pass the standardized filesystem test suite xfstests and that during experimentation and running of the test suite a set of 12 bugs occur. The proportion of memory related bugs found in this set of bugs, is found to be half that of other systems programming projects. Indicating that Rust does provide a positive effect on the memory safety of filesystem implementations.
\pangraph{Future work}
The exact use of the various features of Rust could be further analyzed to research the advantages and disadvantages of the Rust language for filesystem development and potentially expose performance gains or penalties that Rust brings. The latency experiments could also be expanded upon by performing benchmarking on an empty filesystem implementation using the Bento framework, to show whether the framework adds any significant overhead. The \textit{git clone} experiment could be made reproducible by locally hosting a git server on the experimentation machine.

\newpage
\section*{Acknowledgments}
I would like to thank Animesh Trivedi (VU Amsterdam) for supervising me during this research project, and Samantha Miller (University of Washington) and her team who work on Bento for their great work and helping me with issues over email.

\section*{Availability}\label{sec:availability}
The results from this research paper have been made as reproducible as possible by creating a complete benchmarking and data processing suite. This together with the results and crash logs can be found at
\url{https://github.com/Peter-JanGootzen/bento_reproducibility_research}. This should allow anyone to relatively simply reproduce the results from this paper.

\bibliographystyle{plain}
\bibliography{main}

\begin{table*}[b]
	\begin{minipage}[t]{1.0\textwidth}
        \begin{adjustbox}{width=\columnwidth,center}
		    \centering
    		\begin{tabular}{|l|r|r|r|r|}
                \hline
                              \textbf{Workload} & \textbf{ext4-ordered} & \textbf{ext4-journal} & \textbf{Bento-fs} & \textbf{Bento-user} \\\hline
      seq\_read\_4k, T=1 &     2713.05 (193.03) &   2733.16 (245.29) &  2950.72 (320.69) &   3493.6 (364.06) \\
     seq\_read\_32k, T=1 &     4381.29 (357.42) &   4093.39 (387.62) &  5369.05 (676.14) &  4694.15 (444.98) \\
    seq\_read\_128k, T=1 &     4211.15 (479.51) &    4449.2 (303.18) &  4959.29 (154.53) &   5077.1 (111.12) \\
   seq\_read\_1024k, T=1 &     4396.46 (439.99) &   4835.22 (116.77) &  4640.31 (610.74) &  4701.92 (646.84) \\
     seq\_read\_4k, T=18 &     5379.48 (103.83) &    5363.47 (75.32) &  5265.28 (103.97) &   5056.07 (74.46) \\
    seq\_read\_32k, T=18 &    12027.73 (750.65) &  12191.27 (683.54) &  8728.25 (163.41) & 10319.12 (461.54) \\
   seq\_read\_128k, T=18 &    11978.23 (707.59) &  12423.48 (580.52) & 10431.09 (535.49) & 10151.42 (256.98) \\
  seq\_read\_1024k, T=18 &     11760.28 (814.3) &  12230.17 (638.55) & 10112.82 (550.72) &  10010.98 (369.4) \\
     rand\_read\_4k, T=1 &      1911.1 (287.77) &     1364.98 (0.18) &    1365.09 (0.03) &  1774.52 (352.48) \\
    rand\_read\_32k, T=1 &    3071.33 (1079.09) &   2661.83 (989.03) &  2661.88 (989.04) &  3071.3 (1079.07) \\
   rand\_read\_128k, T=1 &       3480.9 (989.0) &     4095.16 (0.07) &  4095.06 (989.01) &  3685.63 (863.32) \\
  rand\_read\_1024k, T=1 &       4096.04 (0.08) &   3891.28 (647.63) &    4096.01 (0.11) &    4096.04 (0.11) \\
    rand\_read\_4k, T=18 &         2047.7 (0.0) &       2047.7 (0.0) &    2047.68 (0.04) &    2047.67 (0.05) \\
   rand\_read\_32k, T=18 &       4095.47 (0.17) &     4095.42 (0.19) &    4095.45 (0.17) &    4095.04 (1.21) \\
  rand\_read\_128k, T=18 &       4097.23 (0.15) &      4096.7 (1.44) &    4097.13 (0.24) &    4097.11 (0.24) \\
 rand\_read\_1024k, T=18 &       4112.94 (0.41) &     4112.85 (0.28) &    4112.83 (0.31) &    4112.86 (0.35) \\\hline
     seq\_write\_4k, T=1 &          999.8 (0.0) &     439.95 (14.07) &      1333.1 (0.0) &       5.89 (0.03) \\
    seq\_write\_32k, T=1 &     1266.45 (140.54) &     483.25 (26.81) &   1666.3 (471.36) &       6.26 (0.05) \\
   seq\_write\_128k, T=1 &     1299.73 (105.38) &        499.9 (0.0) &      1333.1 (0.0) &       6.29 (0.03) \\
  seq\_write\_1024k, T=1 &       1333.08 (0.04) &       488.8 (23.4) &      1999.6 (0.0) &       6.31 (0.06) \\
    rand\_write\_4k, T=1 &         1023.8 (0.0) &       436.8 (23.5) &  1228.55 (176.22) &      13.76 (0.47) \\
   rand\_write\_32k, T=1 &     2084.07 (226.39) &      625.65 (9.21) &  2363.75 (310.87) &      13.76 (0.08) \\
  rand\_write\_128k, T=1 &     2360.56 (194.63) &      670.68 (6.28) &  2729.28 (411.19) &      13.93 (0.46) \\
 rand\_write\_1024k, T=1 &       2048.11 (0.07) &      604.7 (41.11) &    2048.05 (0.05) &      14.83 (0.35) \\
   rand\_write\_4k, T=18 &        791.8 (57.55) &     339.14 (15.98) &   941.94 (105.72) &       14.0 (0.44) \\
  rand\_write\_32k, T=18 &      1602.25 (66.83) &     576.46 (10.24) &  1973.01 (250.07) &      13.62 (0.43) \\
 rand\_write\_128k, T=18 &      2080.7 (109.35) &      628.27 (9.48) &   2243.62 (208.6) &      13.74 (0.58) \\
rand\_write\_1024k, T=18 &       2056.41 (0.46) &      655.99 (47.2) &    2056.58 (0.06) &      14.57 (0.38) \\\hline
      createfiles, T=1 &  132475.92 (12074.2) & 99089.28 (6914.98) &         124981.02 &           2544.25 \\
     createfiles, T=18 & 105344.08 (12125.06) &  88530.6 (5489.65) &          57686.75 &           2674.23 \\
      deletefiles, T=1 &      27494.13 (0.52) &    27493.96 (1.15) &           27494.4 &           1195.51 \\
     deletefiles, T=18 &      27510.48 (0.99) &    27510.35 (1.45) &          27508.15 &           1250.65 \\\hline
            \end{tabular}
        \end{adjustbox}
		\captionsetup{justification=centering,margin=1cm}
		\caption{Results for all filebench workloads with Bento-fs, Bento-fs mounted through user space, Bento-fs with provenance and ext4 ( \texttt{data=journal}  and \texttt{data=ordered} ). The \textit{createfiles} and \textit{deletefiles} workload results denote the average Ops/s over all runs and the standard deviation of Ops/s, the other workload results denote the average MB/s over all runs and the standard deviation of MB/s.}
		\label{tab:filebench}
	\end{minipage}
\end{table*}

\begin{table*}[b]
	\begin{minipage}[t]{1.0\textwidth}
        \begin{adjustbox}{width=\columnwidth,center}
		\centering
            \begin{tabular}{|l|r|r|r|r|}
            \hline
                          \textbf{Workload} &  \textbf{Bento-fs:ext4-j} &  \textbf{Paper Bento-fs:ext4-j} &  \textbf{Bento-user:ext4-j} &  \textbf{Paper Bento-user:ext4-j} \\\hline
      seq\_read\_4k, T=1 &             1.08 &                   1.01 &               1.28 &                     1.01 \\
     seq\_read\_32k, T=1 &             1.31 &                   1.01 &               1.15 &                     1.00 \\
    seq\_read\_128k, T=1 &             1.11 &                   1.01 &               1.14 &                     1.01 \\
   seq\_read\_1024k, T=1 &             0.96 &                   1.02 &               0.97 &                     0.99 \\
     seq\_read\_4k, T=18 &             0.98 &                   1.00 &               0.94 &                     0.99 \\
    seq\_read\_32k, T=18 &             0.72 &                   0.98 &               0.85 &                     0.96 \\
   seq\_read\_128k, T=18 &             0.84 &                   0.99 &               0.82 &                     1.00 \\
  seq\_read\_1024k, T=18 &             0.83 &                   1.00 &               0.82 &                     1.00 \\
     rand\_read\_4k, T=1 &             1.00 &                   1.01 &               1.30 &                     1.00 \\
    rand\_read\_32k, T=1 &             1.00 &                   1.00 &               1.15 &                     0.99 \\
   rand\_read\_128k, T=1 &             1.00 &                   1.00 &               0.90 &                     0.98 \\
  rand\_read\_1024k, T=1 &             1.05 &                   1.00 &               1.05 &                     0.99 \\
    rand\_read\_4k, T=18 &             1.00 &                   1.02 &               1.00 &                     1.02 \\
   rand\_read\_32k, T=18 &             1.00 &                   1.01 &               1.00 &                     1.02 \\
  rand\_read\_128k, T=18 &             1.00 &                   1.00 &               1.00 &                     1.00 \\
 rand\_read\_1024k, T=18 &             1.00 &                   1.00 &               1.00 &                     1.00 \\\hline
     seq\_write\_4k, T=1 &             3.03 &                   1.46 &               0.01 &                     0.02 \\
    seq\_write\_32k, T=1 &             3.45 &                   2.45 &               0.01 &                     0.01 \\
   seq\_write\_128k, T=1 &             2.67 &                   4.12 &               0.01 &                     0.01 \\
  seq\_write\_1024k, T=1 &             4.09 &                   3.93 &               0.01 &                     0.01 \\
    rand\_write\_4k, T=1 &             2.81 &                   1.16 &               0.03 &                     0.07 \\
   rand\_write\_32k, T=1 &             3.78 &                   2.27 &               0.02 &                     0.03 \\
  rand\_write\_128k, T=1 &             4.07 &                   6.55 &               0.02 &                     0.03 \\
 rand\_write\_1024k, T=1 &             3.39 &                  12.24 &               0.02 &                     0.03 \\
   rand\_write\_4k, T=18 &             2.78 &                   1.15 &               0.04 &                     0.04 \\
  rand\_write\_32k, T=18 &             3.42 &                   4.20 &               0.02 &                     0.03 \\
 rand\_write\_128k, T=18 &             3.57 &                   6.24 &               0.02 &                     0.03 \\
rand\_write\_1024k, T=18 &             3.14 &                   6.50 &               0.02 &                     0.04 \\\hline
      createfiles, T=1 &             1.26 &                   1.41 &               0.03 &                     0.02 \\
     createfiles, T=18 &             0.65 &                   1.05 &               0.03 &                     0.01 \\
      deletefiles, T=1 &             1.00 &                   1.09 &               0.04 &                     0.03 \\
     deletefiles, T=18 &             1.00 &                   0.91 &               0.05 &                     0.01 \\\hline
            \end{tabular}
        \end{adjustbox}
        \captionsetup{justification=centering,margin=1cm}
		\caption{The relative performance of Bento-fs through both kernel and user space compared with Linux ext4 with \texttt{data=journal}, together with these same metrics from the original Bento paper \cite{bento}. Note that in the Bento paper T=40 was used instead of T=18.}
		\label{tab:filebench_cmp}
	\end{minipage}
\end{table*}

\end{document}